\documentclass[preprint]{aastex}
\begin{document}

\title
 {MEASUREMENT OF THE COSMIC MICROWAVE BACKGROUND BISPECTRUM 
 ON THE {\it COBE} DMR SKY MAPS}

\shorttitle
 {CMB bispectrum on the {\it COBE} DMR sky maps}


\author
 {E. Komatsu\altaffilmark{1,2},
  B. D. Wandelt\altaffilmark{3},
  D. N. Spergel\altaffilmark{1,4},
  A. J. Banday\altaffilmark{5},
  and K. M. G\'orski\altaffilmark{6,7}}

\shortauthors{Komatsu et al.}

\altaffiltext{1}{Department of Astrophysical Sciences, Princeton University, 
                 Princeton, NJ 08544, USA}
\altaffiltext{1}{Astronomical Institute, T\^ohoku University, Aoba, 
                 Sendai 980-8578, Japan}
\altaffiltext{3}{Department of Physics, Princeton University, 
                 Princeton, NJ 08544, USA}
\altaffiltext{4}{W. M. Keck Distinguished Visiting Professor, 
                 School of Natural Sciences, Institute for Advanced Study, 
		 NJ 08540, USA} 
\altaffiltext{5}{Max Planck Institut f{\"u}r Astrophysik,
                 Karl Schwarzschild Strasse 1, D-85740 Garching bei 
		 M{\"u}nchen, Germany}
\altaffiltext{6}{European Southern Observatory, 
                 Karl Schwarzschild Strasse 2, D-85740 Garching bei 
		 M{\"u}nchen, Germany}
\altaffiltext{7}{Warsaw University Observatory, 
                 Aleje Ujazdowskie 4, 00-478 Warszawa, Poland}

\begin{abstract}
 We measure the angular bispectrum of the cosmic 
 microwave background (CMB) radiation anisotropy from the 
 {\it COBE} Differential Microwave Radiometer (DMR) four-year sky maps.
 The angular bispectrum is the harmonic transform of the three-point
 correlation function, analogous to the angular power spectrum,
 the harmonic transform of the two-point correlation function.
 First, we study statistical properties of the bispectrum and 
 the {\it normalized} bispectrum.
 We find the latter more useful for statistical analysis;
 the distribution of the normalized bispectrum is very much Gaussian, while 
 the bare bispectrum distribution is highly non-Gaussian.
 Then, we measure 466 modes of the normalized bispectrum, all independent 
 combinations of three-point configurations up to a maximum 
 multipole of 20, the mode corresponding to the DMR beam size.
 By measuring 10 times as many modes as the sum of previous work, 
 we test Gaussianity of the DMR maps.
 We compare the data with the simulated Gaussian realizations, finding 
 no significant detection of the normalized bispectrum on the
 mode-by-mode basis. 
 We also find that the previously reported detection of the normalized
 bispectrum is consistent with a statistical fluctuation.
 By fitting a theoretical prediction to the data for the primary 
 CMB bispectrum, which is motivated by slow-roll inflation,
 we put a weak constraint on a parameter characterizing non-linearity in
 inflation. 
 Simultaneously fitting the foreground bispectra estimated from 
 interstellar dust and synchrotron template maps shows that neither 
 dust nor synchrotron emissions significantly contribute to the bispectrum 
 at high Galactic latitude.
 We conclude that the DMR map is consistent with Gaussianity.
\end{abstract}

\section{INTRODUCTION}
\label{sec:intro}


Why study non-Gaussianity of the cosmic microwave background (CMB) 
radiation anisotropy?

Inflation \citep{Guth81,Sato81,AS82,Lin82} predicts Gaussian primordial 
perturbations in quantum origin \citep{GP82,Haw82,Sta82,BST83}, 
implying that two-point statistics such as the angular power spectrum, $C_l$, 
specify all the statistical properties of the CMB anisotropy.
Inflation has passed several challenging observational tests;
the recent CMB experiments \citep{TOCO99,Boom00,Maxima00} have shown that 
the universe is flat as predicted by inflation with a fluctuation spectrum 
consistent with an adiabatic scale-invariant fluctuation.


Several authors have attempted to measure non-Gaussianity 
in CMB using various statistical techniques 
(e.g., Kogut et al. 1996b); as yet no conclusive detection has been 
reported except for measurement of several 
modes of the normalized CMB bispectrum on the {\it COBE} Differential 
Microwave Radiometer (DMR) sky maps \citep{FMG98,Mag00}.
The existence of non-Gaussianity in the DMR data is controversial.
If the CMB sky were non-Gaussian, this would challenge our simplest 
inflationary model.


The angular bispectrum, $B_{l_1l_2l_3}$, is the harmonic transform of 
the three-point correlation function.
We carefully distinguish the normalized bispectrum, 
$B_{l_1l_2l_3}/\left(C_{l_1}C_{l_2}C_{l_3}\right)^{1/2}$, from 
the bispectrum, $B_{l_1l_2l_3}$.
\citet{FMG98} have measured 9 equilateral ($l_1=l_2=l_3$) modes of the 
normalized bispectrum, 
$B_{l_1l_2l_3}/\left(C_{l_1}C_{l_2}C_{l_3}\right)^{1/2}$,
on the DMR map, claiming detection at $l_1=l_2=l_3=16$.
Their result has been under extensive efforts to confirm 
its significance and origin.
\citet{BT99} claim that a few individual pixels in the DMR map 
are responsible for the most of the signal.
\citet{BZG00} have proposed an eclipse effect by the Earth against 
the {\it COBE} satellite as a possible source of the signal.
\citet{Mag00} has measured other 8 inter-$l$ modes of the normalized 
bispectrum such as $B_{l-1ll+1}/\left(C_{l-1}C_{l}C_{l+1}\right)^{1/2}$,
and claims that scatter of the normalized bispectrum among 8 modes 
is too small to be consistent with Gaussian. 
\citet{SM00} further report measurement of 24 other inter-$l$ modes for
different lags in $l$, and conclude they are consistent with Gaussian.


Hence, until now 41 modes of the normalized bispectrum have been 
measured on the DMR map.
Here, we simply ask: ``how many modes are available in the DMR map for 
the bispectrum?''
The answer is 466, up to a maximum multipole of 20 that corresponds to 
the DMR beam size; thus, it is conceivable that the claimed detection of the 
normalized bispectrum at $l_1=l_2=l_3=16$ would be explained by 
a statistical fluctuation, as 9 modes are expected to have statistical 
significance above 98\% out of 466 independent modes even 
if CMB is exactly Gaussian.
In this paper, we measure 466 modes of the CMB bispectrum on the 
{\it COBE} DMR sky maps, testing the claimed detection of the 
bispectrum and non-Gaussianity.
We take into account the covariance between these modes due to the
Galactic cut, which has not been done in the previous work.


On the theoretical side, several predictions for the CMB bispectrum exist.
Several authors \citep{FRS93,LS93,Gan94} have predicted 
the primary bispectrum (or equivalently three-point correlation
function) on the DMR angular scales from slow-roll inflation models. 
\citet{KS01} have extended the prediction 
down to arcminutes scales using the full radiation transfer function.

In addition to the primary one, secondary sources in the low-redshift 
universe and foreground sources produce the bispectrum through their 
non-linearity. 
\citet{LS93} and \citet{SG99} have calculated the secondary bispectrum 
arising from non-linear evolution of gravitational potential;
\citet{GS99} and \citet{CH00} have calculated the one from the gravitational 
lensing effect coupled with various secondary anisotropy sources.
\citet{KS01} have calculated the foreground 
bispectrum from extragalactic radio and infrared point sources.
While the bispectrum is not the best tool for detecting the signature
of rare highly non-linear events, e.g., textures \citep{PK01},
it is sensitive to weakly non-linear effects.


Having theoretical predictions is a great advantage in extracting
physical information from measurement; one can fit a predicted 
bispectrum to the data so as to constrain parameters in a theory. 
Since the DMR beam size is large enough to minimize contribution 
from the secondary and the extragalactic foreground sources, 
the only relevant source would be the primary one.
In this paper, we fit a theoretical primary bispectrum 
\citep{KS01} to the data.


The Galactic plane contains strong microwave emissions from interstellar 
sources.
The emissions are highly non-Gaussian, and distributed on fairly large 
angular scales.
Unfortunately, predicting the CMB bispectrum from interstellar sources
is very difficult; thus, we excise the galactic plane from the DMR data.
We model the residual foreground bispectrum at high
galactic latitude using foreground template maps.
By simultaneously fitting the foreground bispectrum and the primary 
bispectrum to the DMR data for three different Galactic cuts, 
we quantify the importance of the interstellar emissions in our analysis.


This paper is organized as follows.
In \S~\ref{sec:bispectrum}, we define the angular bispectrum, and 
show how to compute it efficiently from observational data. 
In \S~\ref{sec:measurement}, we study statistical properties of the 
bispectrum and the normalized bispectrum.
We then measure the normalized bispectrum on the {\it COBE} DMR 
four-year sky maps \citep{Ben96}, testing Gaussianity of the DMR map.
In \S~\ref{sec:fit}, we fit predicted bispectra to the DMR data, 
constraining parameters in the predictions. 
The predictions include the primary bispectrum from inflation and the 
foreground bispectrum from interstellar Galactic emissions.
Finally, \S~\ref{sec:discussion_bl} concludes.
In the appendix, we derive the relations between the angular 
power spectrum and 
bispectrum on the incomplete sky and those on the full sky.

\section{ANGULAR BISPECTRUM}
\label{sec:bispectrum}

The CMB angular bispectrum consists of a product of three 
harmonic transforms of the CMB temperature field.
For Gaussian fields, expectation value of the bispectrum is exactly zero.
Given statistical isotropy of the universe, the angular averaged 
bispectrum, $B_{l_1l_2l_3}$, is given by
\begin{equation}
  \label{eq:blll}
  B_{l_1l_2l_3}= \sum_{{\rm all}~m}
  \left(
  \begin{array}{ccc}
  l_1&l_2&l_3\\
  m_1&m_2&m_3
  \end{array}
  \right)
  a_{l_1m_1}a_{l_2m_2}a_{l_3m_3},
\end{equation}
where the matrix denotes the Wigner-$3j$ symbol.
The harmonic coefficients, $a_{lm}$, are given by
\begin{equation}
 \label{eq:alm}
  a_{lm}= 
  \int_{\Omega_{\rm obs}} 
  d^2\hat{\mathbf n}~\frac{\Delta T\left(\hat{\mathbf n}\right)}T
  Y_{lm}^*\left(\hat{\mathbf n}\right),
\end{equation}
where $\Omega_{\rm obs}$ denotes a solid angle of the observed sky. 
$B_{l_1l_2l_3}$ satisfies the triangle condition,
$\left|l_i-l_j\right|\leq l_k \leq l_i+l_j$ for all permutations of
indices, and parity invariance, $l_1+l_2+l_3={\rm even}$.

We rewrite equation~(\ref{eq:blll}) into a more computationally
efficient form.
Using the identity,
\begin{eqnarray}
 \nonumber
  \left(
   \begin{array}{ccc}
    l_1 & l_2 & l_3 \\ m_1 & m_2 & m_3 
   \end{array}
 \right)
  &=&
  \left(
   \begin{array}{ccc}
    l_1 & l_2 & l_3 \\ 0 & 0 & 0 
   \end{array}
 \right)^{-1}
  \sqrt{
  \frac{(4\pi)^3}{\left(2l_1+1\right)\left(2l_2+1\right)\left(2l_3+1\right)}
  }\\
 \label{eq:gaunt}
 & &\times
  \int \frac{d^2\hat{\mathbf n}}{4\pi}~
  Y_{l_1m_1}(\hat{\mathbf n})
  Y_{l_2m_2}(\hat{\mathbf n})
  Y_{l_3m_3}(\hat{\mathbf n}),
\end{eqnarray}
we rewrite equation~(\ref{eq:blll}) as
\begin{equation}
  \label{eq:bobs}
  B_{l_1l_2l_3}=
  \left(
   \begin{array}{ccc}
    l_1 & l_2 & l_3 \\ 0 & 0 & 0 
   \end{array}
 \right)^{-1}
  \int \frac{d^2\hat{\mathbf n}}{4\pi}~
  e_{l_1}(\hat{\mathbf n})
  e_{l_2}(\hat{\mathbf n})
  e_{l_3}(\hat{\mathbf n}),
\end{equation}
where the integral is not over $\Omega_{\rm obs}$, but over the 
whole sky; $e_l(\hat{\mathbf n})$ already encapsulates 
the information of incomplete sky coverage through $a_{lm}$.
Here, following \citet{SG99}, we have used the azimuthally averaged harmonic 
transform of the CMB temperature field, $e_l(\hat{\mathbf n})$,
\begin{equation}
  \label{eq:el}
   e_{l}(\hat{\mathbf n})
   =
   \sqrt{\frac{4\pi}{2l+1}}
   \sum_m a_{lm} Y_{lm}(\hat{\mathbf n}).
\end{equation}
Similarly, we write the angular power spectrum, $C_l$, as
\begin{equation}
  \label{eq:cl}
   C_l=   
   \int \frac{d^2\hat{\mathbf n}}{4\pi}~
   e^2_{l}(\hat{\mathbf n}).
\end{equation}
$e_l(\hat{\mathbf n})$ is thus a square-root of $C_l$ at a given 
position of the sky.

Equation~(\ref{eq:bobs}) is computationally efficient, as we can calculate 
$e_l(\hat{\mathbf n})$ quickly with the spherical harmonic transform 
for a given $l$. 
Since the HEALPix pixels have the equal area \citep{GHW98},
the average over the whole sky, $\int d^2\hat{\mathbf n}/(4\pi)$,
is done by the sum over all pixels divided by the total number 
of pixels, $N^{-1}\sum_i^{N}$.

\section{MEASUREMENT OF BISPECTRUM ON THE {\it COBE} DMR SKY MAPS}
\label{sec:measurement} 

\subsection{The data}

We use the HEALPix-formatted \citep{GHW98} {\it COBE} DMR four-year sky
map, which contains 12,288 pixels in Galactic coordinate
with a pixel size $1.\hspace{-4pt}^\circ 83$.
We obtain the most sensitive sky map to CMB by combining 53~GHz map with 
90~GHz map, after coadding the channels A and B at each frequency.
We do not subtract eclipse season time-ordered data;
while \citet{BZG00} ascribe the reported non-Gaussianity to this data,
we will argue in this paper that the claimed detection of the normalized
bispectrum at $l_1=l_2=l_3=16$ \citep{FMG98} can also be explained 
in terms of a statistical fluctuation.

We reduce interstellar Galactic emissions by using three different
Galactic cuts: the $20^\circ$ cut, the extended cut \citep{Ban97},
and the $25^\circ$ cut.
We then subtract the monopole and the dipole from each cut map,
minimizing contaminations from these two multipoles to
higher order multipoles through the mode-mode coupling.
The coupling arises from incomplete sky coverage.
This is very important to do, for the leakage of power from 
the monopole and the dipole to the higher order multipoles is rather big.
We use the least-squares fit weighted by the pixel noise variance
to measure the monopole and the dipole on each cut map.

We measure the bispectrum, $B_{l_1l_2l_3}$, on the DMR sky maps as follows.
First, we measure $a_{lm}$ using equation~(\ref{eq:alm}).
Then, we transform $a_{lm}$ for $-l\le m\le l$ into 
$e_l(\hat{\mathbf n})$ through equation~(\ref{eq:el}).
Finally, we obtain $B_{l_1l_2l_3}$ from equation~(\ref{eq:bobs}),
arranging $l_1$, $l_2$, and $l_3$ in order of $l_1\le l_2\le l_3$, where
we set the maximum $l_3$ to be 20.
In total, we have 466 non-zero modes after taking into account
$\left|l_i-l_j\right|\leq l_k \leq l_i+l_j$ and $l_1+l_2+l_3={\rm even}$.
Measurement of 466 modes takes about 1 second of CPU time on 
a Pentium-III single processor personal computer.

\subsection{Monte--Carlo Simulations}

We use Monte--Carlo simulations to estimate the covariance matrix of the 
measured bispectrum.
Our simulation includes 
(a) a Gaussian random realization of the primary CMB anisotropy field 
drawn from the {\it COBE}-normalized ${\Lambda}$CDM power spectrum,
and (b) a Gaussian random realization of the instrumental noise drawn from
diagonal terms of the {\it COBE} DMR noise covariance matrix \citep{Lin94}.
For computational efficiency, we do not use off-diagonal terms 
as they are smaller than 1\% of the diagonal terms \citep{Lin94}.

We generate the input power spectrum, $C_l$, using the {\sf CMBFAST} code 
\citep{SZ96} with cosmological parameters fixed at 
$\Omega_{\rm cdm}=0.25$, $\Omega_\Lambda=0.7$, $\Omega_{\rm b}=0.05$, 
$h=0.7$, and $n=1$; the {\sf CMBFAST} code uses the \citet{BW97} 
power-spectrum normalization.

In each realization, we generate $a_{lm}$ from the power spectrum, 
multiply it by the harmonic-transformed DMR beam, $G_l$
\citep{Wri94}, transform $G_l a_{lm}$ back to a sky map, and add an
instrumental noise realization to the map.
Finally, we measure 466 modes of the bispectrum from each realization. 
We generate 50,000 realizations for one simulation;
processing one realization takes about 1 second, so that 
one simulation takes about 16 hours of CPU time on a Pentium-III single 
processor personal computer.

\subsection{Normalized bispectrum}

The input power spectrum determines the variance of the bispectrum.
Off-diagonal terms in the covariance matrix arise from incomplete 
sky coverage.
When non-Gaussianity is weak, the variance is given by 
\citep{Luo94,heavens98,SG99,GM00}
\begin{equation}
 \label{eq:var}
  \left<B_{l_1l_2l_3}^2\right>
  = \left<C_{l_1}\right>\left<C_{l_2}\right>\left<C_{l_3}\right>
  \Delta_{l_1l_2l_3},
\end{equation}
where $\Delta_{l_1l_2l_3}$ takes values 1, 2, or 6 
for all $l$'s are different, two are same, or all are same, respectively.
The brackets denote the ensemble average.

The variance is undesirably sensitive to the input power spectrum;
even if the input power spectrum were slightly different 
from the true power spectrum on the DMR map, the estimated variance 
from simulations would be significantly wrong,
and we would erroneously conclude that the DMR map is inconsistent 
with Gaussian.
It is thus not a robust test of Gaussianity to  
compare the measured bispectrum with the Monte--Carlo simulations.

The {\it normalized} bispectrum, 
$B_{l_1l_2l_3}/\left(C_{l_1}C_{l_2}C_{l_3}\right)^{1/2}$,
is more sensible quantity than the bare bispectrum.
\citet{Mag95} shows that the normalized bispectrum is 
a rotationally invariant spectrum independent of the power spectrum,
as it factors out fluctuation amplitude in $a_{lm}$, which is 
measured by $C_l^{1/2}$.
By construction, the variance of the normalized bispectrum is 
insensitive to the power spectrum, approximately given by 
$\Delta_{l_1l_2l_3}$.

\begin{figure}
 \plotone{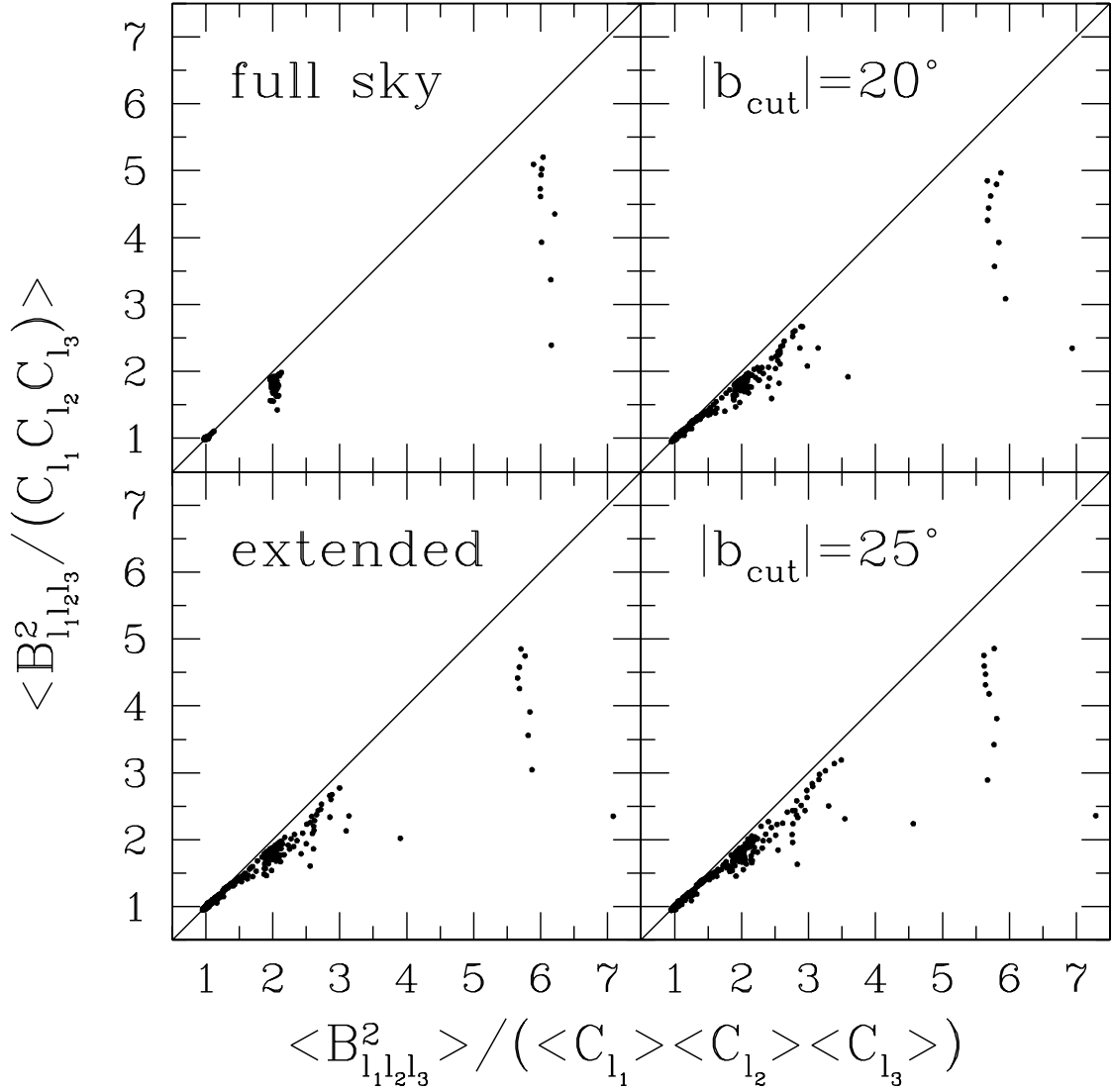}
 \caption
 {Comparison of the variance of the normalized bispectrum, 
 $\left<B_{l_1l_2l_3}^2/\left(C_{l_1}C_{l_2}C_{l_3}\right)\right>$,
 with that of the bare bispectrum,
 $\left<B_{l_1l_2l_3}^2\right>/
 \left(\left<C_{l_1}\right>\left<C_{l_2}\right>\left<C_{l_3}\right>\right)$.
 The top-left panel shows the case of full sky coverage, while
 the rest of panels show the cases of incomplete sky coverage.
 The top-right, bottom-left, and bottom-right panels use
 the $20^\circ$ cut, the extended cut, and the $25^\circ$
 cut, respectively.}
\label{fig:variance}
\end{figure}

One might wonder if the normalized bispectrum is too noisy to be
useful, as the power spectrum in the denominator is also uncertain; 
however, we find that the variance is actually
slightly smaller than $\Delta_{l_1l_2l_3}$.
Figure~\ref{fig:variance} compares the variance of the normalized
bispectrum, 
$\left<B_{l_1l_2l_3}^2/\left(C_{l_1}C_{l_2}C_{l_3}\right)\right>$,
with that of the bispectrum,
$\left<B_{l_1l_2l_3}^2\right>/
\left(\left<C_{l_1}\right>\left<C_{l_2}\right>\left<C_{l_3}\right>\right)$.
The top-left panel shows the case of full sky coverage. 
We find that the variance of the normalized bispectrum is precisely 1 
when all $l$'s are different, while it is slightly smaller than 2 or 6 
when two $l$'s are same or all $l$'s are same, respectively.
This arises due to correlation between the uncertainties in the 
bispectrum and the power spectrum, and this correlation tends to reduce 
the total variance of the normalized bispectrum.
The rest of panels show the cases of incomplete sky coverage.
While the variance becomes more scattered than the case of 
full sky coverage, the variance of the normalized bispectrum is 
still systematically smaller than that of the bare bispectrum.
The normalized bispectrum is thus reasonably sensitive to
non-Gaussianity, yet it is not sensitive to the overall normalization
of power spectrum.

\begin{figure}
 \plotone{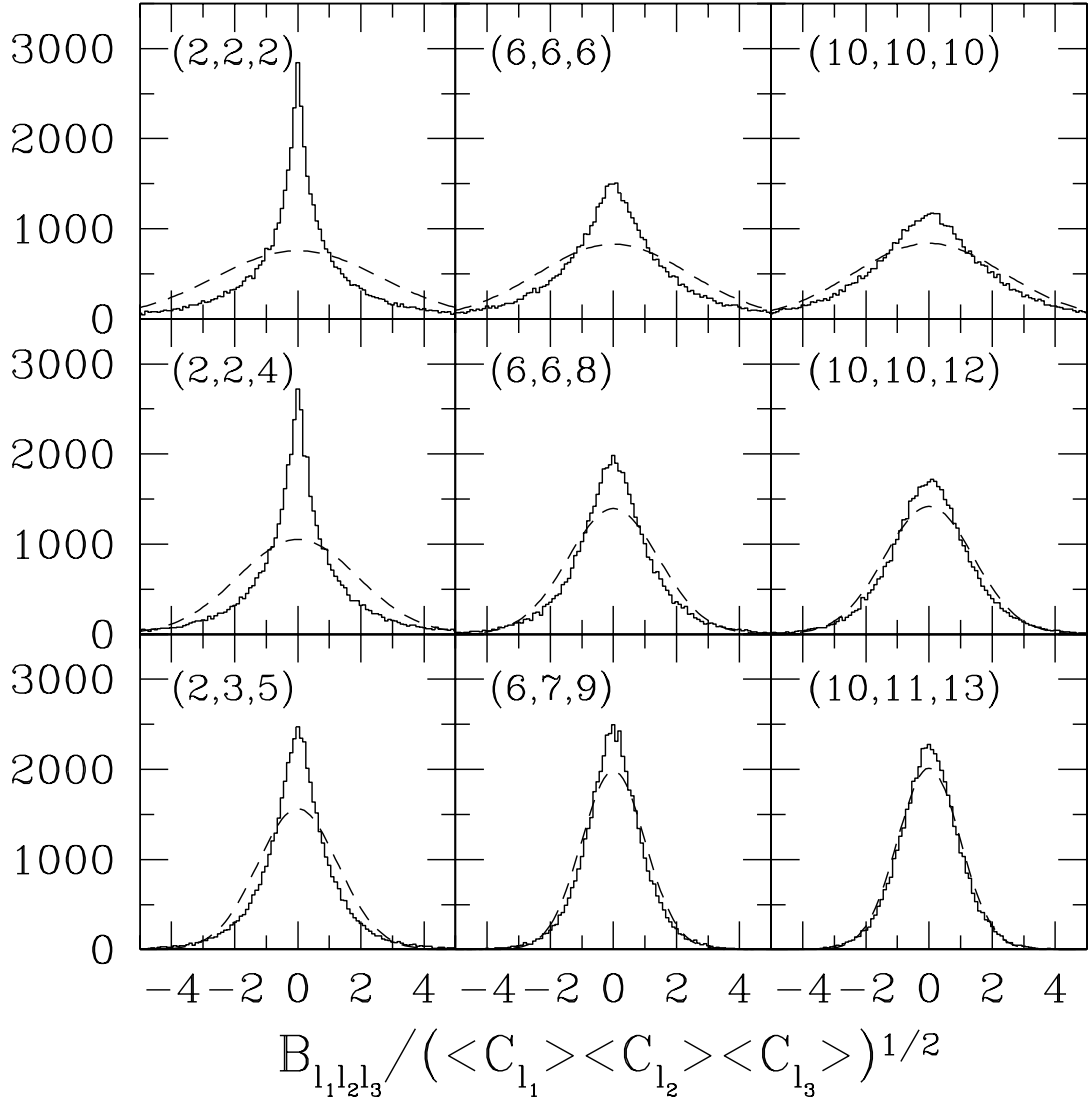}
 \caption
 {Distribution of the bispectrum drawn from the Monte--Carlo simulations
 for the $20^\circ$ Galactic cut (solid lines).
 $B_{l_1l_2l_3}/\left(\left<C_{l_1}\right>
 \left<C_{l_2}\right>\left<C_{l_3}\right>\right)^{1/2}$ is plotted,
 where the brackets denote the ensemble average over realizations
 from the Monte--Carlo simulations.
 The dashed lines plot Gaussian distributions calculated from 
 r.m.s. values.
 Each panel represents a certain mode of $(l_1,l_2,l_3)$ as quoted in 
 the panels.}
\label{fig:dist_bare}
\end{figure}

\begin{figure}
 \plotone{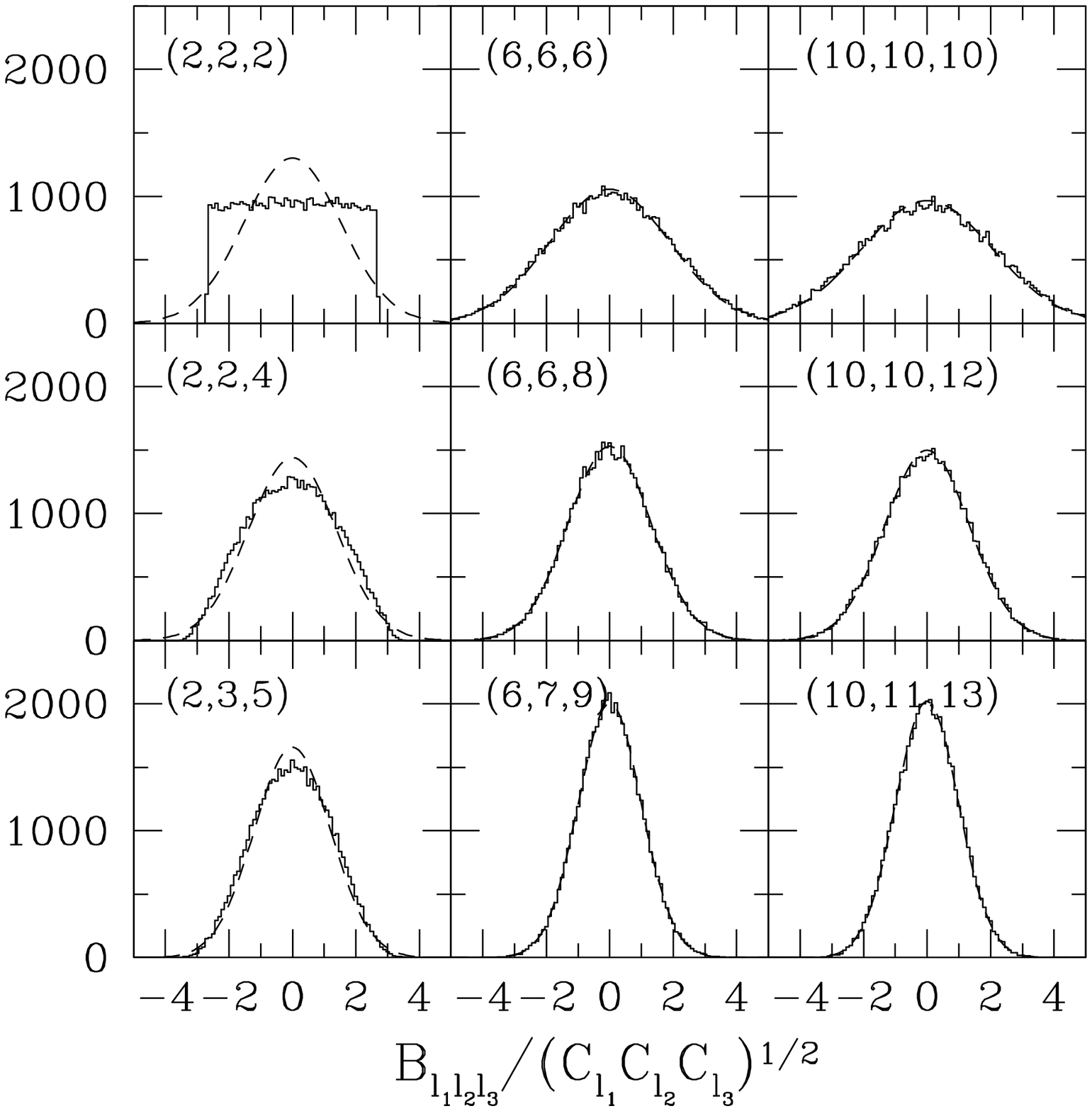}
 \caption
 {Distribution of the normalized bispectrum,
 $B_{l_1l_2l_3}/\left(C_{l_1}C_{l_2}C_{l_3}\right)^{1/2}$,
 drawn from the Monte--Carlo simulations for the $20^\circ$ Galactic cut.
 The meaning of the lines is the same as in figure~\ref{fig:dist_bare}.}
\label{fig:dist_norm}
\end{figure}

What distribution does the normalized bispectrum obey for a Gaussian field?
First, even for a Gaussian field, the probability distribution 
of a single mode of $B_{l_1l_2l_3}$ is non-Gaussian,
characterized by a large kurtosis.
Figure~\ref{fig:dist_bare} plots the distributions of 9 modes of  
$B_{l_1l_2l_3}$ drawn from the Monte--Carlo simulations 
(solid lines) in comparison with Gaussian distributions calculated from 
r.m.s. values (dashed lines).
We find that the distribution does not fit the Gaussian very well.
Then, we examine distribution of the normalized bispectrum, 
$B_{l_1l_2l_3}/\left(C_{l_1}C_{l_2}C_{l_3}\right)^{1/2}$.
We find that the distribution is very much Gaussian except 
for $l_1=l_2=l_3=2$.
Figure~\ref{fig:dist_norm} plots the distributions of the 9 modes 
of the normalized bispectrum (solid lines) in comparison with 
Gaussian distributions calculated from r.m.s. values (dashed lines).
The distribution fits the Gaussian remarkably well; this
motivates our using standard statistical methods developed for 
Gaussian fields to analyze the normalized bispectrum.
We could not make this simplification if we were analyzing the bare bispectrum.
Furthermore, the central limit theorem implies that when we combine 466 modes
the deviation of the distribution from Gaussianity becomes even smaller.

\citet{FMG98} claim detection of the normalized bispectrum 
at $l_1=l_2=l_3=16$; 
\citet{Mag00} claims that the scatter of the 
normalized bispectrum for $l_1=l_2-1$ and $l_3=l_2+1$ is 
too small to be consistent with Gaussian.
The former has analyzed 9 modes, while the latter has analyzed 8 modes.
In the next section, we analyze 466 modes, testing the statistical 
significance of the non-Gaussianity with much more samples than the 
previous work.
We calculate $C_l$ from equation~(\ref{eq:cl}), and then
divide $B_{l_1l_2l_3}$ by $\left(C_{l_1}C_{l_2}C_{l_3}\right)^{1/2}$
to obtain the normalized bispectrum.

\subsection{Testing Gaussianity of the DMR map}

We characterize statistical significance of the normalized bispectrum
as probability of the measured normalized bispectrum being greater than 
those drawn from the Monte--Carlo simulations.
We define the probability $P$ as 
\begin{equation}
 \label{eq:significance}
  P_\alpha\equiv 
  \frac{N\left(\left|I_{\alpha}^{\rm DMR}\right|
	  >\left|I_{\alpha}^{\rm MC}\right|\right)}
  {N_{\rm total}}
  =
  \int_{-\left|I_\alpha^{\rm DMR}\right|}^{\left|I_\alpha^{\rm DMR}\right|}
  dx~F_\alpha^{\rm MC}(x),
\end{equation}
where $I_\alpha$ is the normalized bispectrum,
$N_{\rm total}=50,000$ is the total number of simulated realizations,
and $\alpha=$ 1, 2, 3, 4,\dots, 466 represent
$(l_1,l_2,l_3)=$
(2,2,2), (2,3,3), (2,2,4), (3,3,4),\dots, (20,20,20), respectively,
with satisfying $l_1\le l_2\le l_3$, 
$\left|l_i-l_j\right|\leq l_k \leq l_i+l_j$, and $l_1+l_2+l_3={\rm even}$.
$F_\alpha^{\rm MC}(x)$ is the p.d.f of the simulated 
realizations for the normalized bispectrum, $x=I_{\alpha}^{\rm MC}$.
The p.d.f is normalized to unity: 
$\int_{-\infty}^\infty dx F_\alpha^{\rm MC}(x) =1$; thus,
$P_\alpha$ lies in $0\le P_\alpha\le 1$.

By construction, the distribution of $P_\alpha$ is uniform, if
the DMR map is consistent with the simulated realizations, i.e.,
Gaussian.
We give the proof as follows.
By rewriting equation~(\ref{eq:significance}) as 
$P_\alpha=f(\left|I_\alpha^{\rm DMR}\right|)$, we calculate the p.d.f
of $P_\alpha$, $G(P_\alpha)$, as
\begin{eqnarray}
 \nonumber
 G(P_\alpha)&=& 
  \int_{-\infty}^\infty dy~\delta\left[P_\alpha=f(\left|y\right|)\right]
  F_\alpha^{\rm DMR}(y)\\
  &=&
   \nonumber
   \int_{0}^\infty dy~\delta\left[P_\alpha=f(y)\right]
   \left[F_\alpha^{\rm DMR}(y)+F_\alpha^{\rm DMR}(-y)\right]\\
 &=&
  \nonumber
   \int_{0}^\infty dy~
   \frac{\delta\left[y=f^{-1}(P_\alpha)\right]}{\left|df/dy\right|}
   \left[F_\alpha^{\rm DMR}(y)+F_\alpha^{\rm DMR}(-y)\right]\\
 &=&
  \label{eq:uniformity}
  \int_{0}^\infty dy~\delta\left[y=f^{-1}(P_\alpha)\right]
  \frac{F_\alpha^{\rm DMR}(y)+F_\alpha^{\rm DMR}(-y)}
  {F^{\rm MC}_\alpha(y)+F^{\rm MC}_\alpha(-y)},
\end{eqnarray}
where $F_\alpha^{\rm DMR}(y)$ is the p.d.f of
the measured normalized bispectrum on the DMR map, $y=I_\alpha^{\rm DMR}$.
Our goal is to see if $F_\alpha^{\rm DMR}(y)$ is consistent with
the DMR data being Gaussian.
It follows from equation~(\ref{eq:uniformity}) 
that $G(P_\alpha)\equiv 1$, when 
$F_\alpha^{\rm DMR}(y)\equiv F^{\rm MC}_\alpha(y)$, regardless of 
the functional form of $F^{\rm MC}_\alpha(y)$.
In other words, the distribution of $P_\alpha$ is uniform, if
the distribution of the measured normalized bispectrum is the same as
the simulated realizations.
Since our simulation assumes the DMR map Gaussian, the $P$ 
distribution, $G(P)$, tests Gaussianity of the DMR map.
If the $P$ distribution is not uniform, then we conclude the DMR data 
to be non-Gaussian.

A Gaussian field gives equal number of modes 
in each bin of $P$. 
For example, it gives 46.6 modes in $\Delta P=10\%$ bin:
$466\times G(P)\Delta P=466\times 0.1= 46.6$.
If we detect the normalized bispectrum significantly, then we find that
$G(P)$ is not uniform, but increases rapidly as $P$ increases.

\begin{figure}
 \plotone{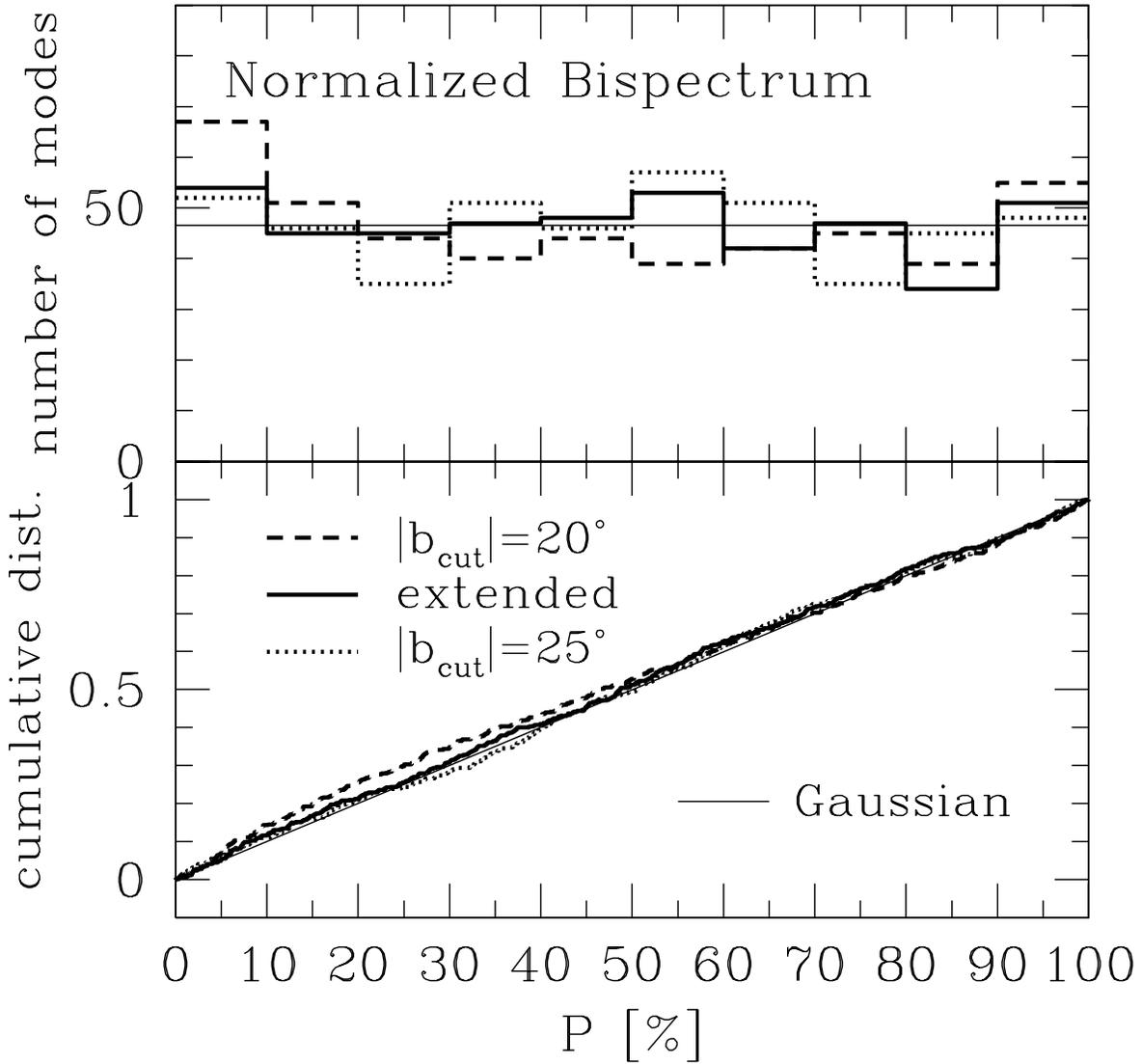}
 \caption
 {$P$ distribution (Eq.(\ref{eq:significance})).
 $P$ is the probability of the CMB normalized bispectrum, 
 $B_{l_1l_2l_3}/\left(C_{l_1}C_{l_2}C_{l_3}\right)^{1/2}$, 
 measured on the {\it COBE} DMR $53+90~{\rm GHz}$ sky map, 
 being larger than those drawn from the Monte--Carlo simulations.
 There are 466 modes in total.
 The thick dashed, solid, and dotted lines represent 
 the three different Galactic cuts as quoted in the figure.
 The thin solid line shows the expectation value for a Gaussian field.
 The top panel shows the $P$ distribution, while the bottom panel
 shows the cumulative $P$ distribution,
 for which we calculate the KS statistic.
 The KS statistic gives the probability of the distribution being 
 consistent with the expectation for Gaussianity as
 6.7\%, 73\%, and 77\% for the three Galactic cuts, respectively.}
\label{fig:KS_norm}
\end{figure}

The top panel of figure~\ref{fig:KS_norm} plots the $P$ distribution
for the three different Galactic cuts.
We find that the distribution is uniform, and the number of modes
in the bin ($\Delta P=10\%$) is consistent with the expectation value 
for Gaussian fluctuations (46.6).

To further quantify how well it is uniform, we calculate 
the Kolmogorov--Smirnov (KS) statistic for the $P$ distribution in 
comparison with the uniform distribution.
The bottom panel of figure~\ref{fig:KS_norm} plots the cumulative 
$P$ distribution, for which we calculate the KS statistic.
The probability of the distribution being uniform is 6.7\%, 73\%, and 77\% 
for the three Galactic cuts, respectively.

We have confirmed that the normalized bispectrum at $l_1=l_2=l_3=16$ 
has $P=97.81\%$ for the $20^\circ$ cut, $P=99.97\%$ for the extended
cut, and $P=99.27\%$ for the $25^\circ$ cut, as similar to 
\citet{FMG98}; however, our result shows that statistical fluctuations 
explain the significance.
We conclude that the properties of the normalized bispectrum of the 
DMR map are consistent with CMB being a Gaussian field.

\section{MODEL FITTING}
\label{sec:fit}

In this section, we fit predicted CMB bispectra to the measured 
normalized bispectrum.
The predictions include the primary bispectrum from inflation
and the interstellar foreground bispectrum from the Galactic emissions.
Then, we constrain a parameter characterizing the primary bispectrum.

\subsection{Primary bispectrum}

For the primary bispectrum from inflation, we consider weakly 
non-Gaussian adiabatic perturbations generated through non-linearity 
in slow-roll inflation. 
The simplest weak non-linear coupling gives
\begin{equation}
  \label{eq:modelreal}
  \Phi({\mathbf x})
 =\Phi_{\rm L}({\mathbf x})
 +f_{\rm NL}\left[
              \Phi^2_{\rm L}({\mathbf x})-
	      \left<\Phi^2_{\rm L}({\mathbf x})\right>
        \right],
\end{equation}
where the square bracket denotes the volume average, and 
$\Phi_{\rm L}({\mathbf x})$ is a linear Gaussian part of 
curvature perturbations. 
We call $f_{\rm NL}$ the non-linear coupling parameter, following 
\citet{KS01}.

\citet{SB90,SB91} and \citet{Gan94} show that
slow-roll inflation gives this coupling; 
\citet{PC96} shows that the second-order general relativistic 
perturbation theory gives this.  
The former predicts $f_{\rm NL}$ as a certain combination of 
slope and curvature of a inflaton potential
($\Phi_3=-2f_{\rm NL}$ in \citet{Gan94}; $\alpha_\Phi=f_{\rm NL}$
in \citet{VWHK00}).
The latter predicts $f_{\rm NL}\sim{\cal O}(1)$.
\citet{KS01} have given the exact form of 
$B_{l_1l_2l_3}$ for this model; we do not repeat it here.
$f_{\rm NL}$ is the parameter that we try to constrain by measuring
the CMB bispectrum.

Since the theoretical bispectrum assumes full sky coverage,
we must correct it for the bias arising from incomplete sky
coverage.
We use an approximate correction factor for the bias, 
$\Omega_{\rm obs}/4\pi$, which we have derived in the appendix.
Moreover, the theoretical bispectrum must also be convolved with the 
DMR beam.
We use the harmonic transform of the DMR beam, $G_l$, given in \citet{Wri94}.
Hence, we relate the observed bispectrum to the theoretical bispectrum
through
\begin{equation}
 \label{eq:theory2obs}
  B_{l_1l_2l_3}^{\rm obs}
  =\frac{\Omega_{\rm obs}}{4\pi}
  B_{l_1l_2l_3}^{\rm theory}
  G_{l_1}G_{l_2}G_{l_3}.
\end{equation}
Note that $\Omega_{\rm obs}/4\pi= 1-\sin \left|b_{\rm cut}\right|$
for an azimuthally symmetric cut within certain latitude $b_{\rm cut}$;
$\Omega_{\rm obs}/4\pi= 0.658$, 0.577, and 0.5 for 
$\left|b_{\rm cut}\right|=20^\circ$, $25^\circ$, and $30^\circ$, 
respectively.
For the extended cut, $\Omega_{\rm obs}/4\pi= 0.638$.

\subsection{Foreground bispectra from interstellar emissions}

Although we cut a fraction of the sky to reduce interstellar 
emissions from the Galactic plane, there should be some residuals
at high Galactic latitude.
\citet{Kog96a} have found significant correlation between {\it COBE} DMR
maps at high Galactic latitude and {\it COBE} Diffuse Infrared 
Background Experiment (DIRBE) map which mainly trace dust emission 
from the Galactic plane.

The interstellar emissions are highly non-Gaussian.
For example, the one-point p.d.f of the all-sky dust template map 
\citep{SFD98} is highly skewed.
We find the normalized skewness, 
$\left<(\Delta T)^3\right>/\left<(\Delta T)^2\right>^{3/2}\sim 51$.
Since these non-Gaussian emissions would confuse the parameter 
estimation of the primary CMB bispectrum, we take the effect into account.

We estimate the foreground bispectra from interstellar sources
by using two foreground template maps.
One is the dust template map of \citet{SFD98}; the other is the 
synchrotron map of \citet{Has81}.
Both maps are in the HEALPix format \citep{GHW98}.

We extrapolate the dust map to 53~GHz and 90~GHz with taking into account
spatial variations of dust temperatures across the sky \citep{Fin99}.
We then cross-correlate the extrapolated maps with the DMR maps
to confirm that the extrapolation is reasonable.
We find that while the dust-correlated emission in the DMR 90~GHz map 
is consistent with the extrapolated dust emission,
that in the DMR 53~GHz map is much larger than the extrapolated one.
This is consistent with the anomalous microwave emission of \citet{Kog96a}. 
To take the excess emission into account, we multiply our
extrapolated 53~GHz maps by factors of 3.66, 2.58, and 2.59
for the $20^\circ$ cut, the extended cut, and the $25^\circ$
cut, respectively.
The correction factor is notably larger for the $20^\circ$ cut
rather than the extended or $25^\circ$ cut .
This could possibly be attributed
to the region around Ophiucus which has a free-free spectrum.
Note that we do not essentially need the correction for the excess emission,
as it does not alter spatial distribution of the emission.
Nevertheless, we do it for convenience of subsequent analyses.

We also extrapolate the synchrotron map to these two bands, assuming
the spectrum of the source, $T(\nu)\propto \nu^{-2.9}$. 
We do not need the extrapolation of the synchrotron template map either, 
as the extrapolation does not alter spatial distribution of 
the emission in contrast to the dust template map in which the 
extrapolation does alter it.
We find no significant correlation between the DMR maps and the 
extrapolated synchrotron maps at both 53~GHz and 90~GHz.

After coadding the extrapolated 53 and 90~GHz maps with the same 
weight as used for the DMR maps, we measure $B_{l_1l_2l_3}$ from the
maps for the three different Galactic cuts, multiplying it by 
$G_{l_1}G_{l_2}G_{l_3}$ to take into account the DMR beam.

\subsection{Constraints on non-linearity in inflation}

We fit simultaneously the primary, dust, and synchrotron bispectra to 
the measured bispectrum on the DMR map.
We use the least-squares method based on a $\chi^2$ statistic defined by
\begin{equation}
  \label{eq:chisq}
  \chi^2(f_j)\equiv 
  \sum_{\alpha\alpha'}
  \left(I^{\rm DMR}_\alpha-\sum_jf_jI_\alpha^{j}\right)
  \left(C^{-1}\right)_{\alpha\alpha'}
  \left(I^{\rm DMR}_{\alpha'}-\sum_jf_jI_{\alpha'}^{j}\right).
\end{equation}
$I_\alpha^{j}$ is a model bispectrum divided by 
$\left(\left<C_{l_1}^{\rm MC}\right>
\left<C_{l_2}^{\rm MC}\right>\left<C_{l_3}^{\rm MC}\right>\right)^{1/2}$, 
where $j$ represents a certain component such as the primary, dust, 
and synchrotron. 
$f_j$ is a fitting parameter for a component $j$,
where $f_{\rm primary}\equiv f_{\rm NL}$ is the non-linear coupling
parameter (Eq.(\ref{eq:modelreal})). 
$f_{\rm dust}$ and $f_{\rm sync}$ characterize amplitude of the 
foreground bispectra.

$C_{\alpha\alpha'}$ is the covariance matrix of the 
normalized bispectrum, which we calculate from the Monte--Carlo simulations:
\begin{equation}
  \label{eq:cov}
  C_{\alpha\alpha'}\equiv
  \frac1{N-1} 
  \sum_{i=1}^{N}
  \left(I^{{\rm MC}(i)}_\alpha -\left<I^{\rm MC}_\alpha \right>\right)
  \left(I^{{\rm MC}(i)}_{\alpha'}-\left<I^{\rm MC}_{\alpha'}\right>\right),
\end{equation}
where $N=50,000$ is the total number of realizations. 
The bracket denotes an average over all realizations,
$\left<I^{\rm MC}_\alpha \right>\equiv N^{-1} \sum_i^N 
I_\alpha^{{\rm MC}(i)}$.  
Here, we have implicitly assumed the non-Gaussianity weak, so that
we calculate the covariance matrix from Gaussian realizations.

As we have observed in the previous section, distribution of
$I_\alpha$ is very much Gaussian; thus, $\chi^2(f_j)$ should obey 
the $\chi^2$ distribution to good accuracy.
Hence, minimizing $\chi^2(f_j)$ with respect to $f_j$ gives 
the maximum-likelihood value of $f_j$ as a solution to the normal equation,
\begin{equation}
 \label{eq:normal}
  f_j = \sum_i\left(F^{-1}\right)_{ji}
  \left[\sum_{\alpha\alpha'}
  I_\alpha^i\left(C^{-1}\right)_{\alpha\alpha'}I_{\alpha'}^{\rm DMR}
 \right],
\end{equation}
where
\begin{equation}
 \label{eq:fisher}
  F_{ij}\equiv
  \sum_{\alpha\alpha'}
  I_\alpha^i\left(C^{-1}\right)_{\alpha\alpha'}I_{\alpha'}^j.
\end{equation}
We estimate statistical uncertainties of the parameters using 
the Monte--Carlo simulations; we obtain parameter realizations
by substituting $I_\alpha^{\rm MC}$ for $I_\alpha^{\rm DMR}$ 
in equation~(\ref{eq:normal}).

\begin{figure}
 \plotone{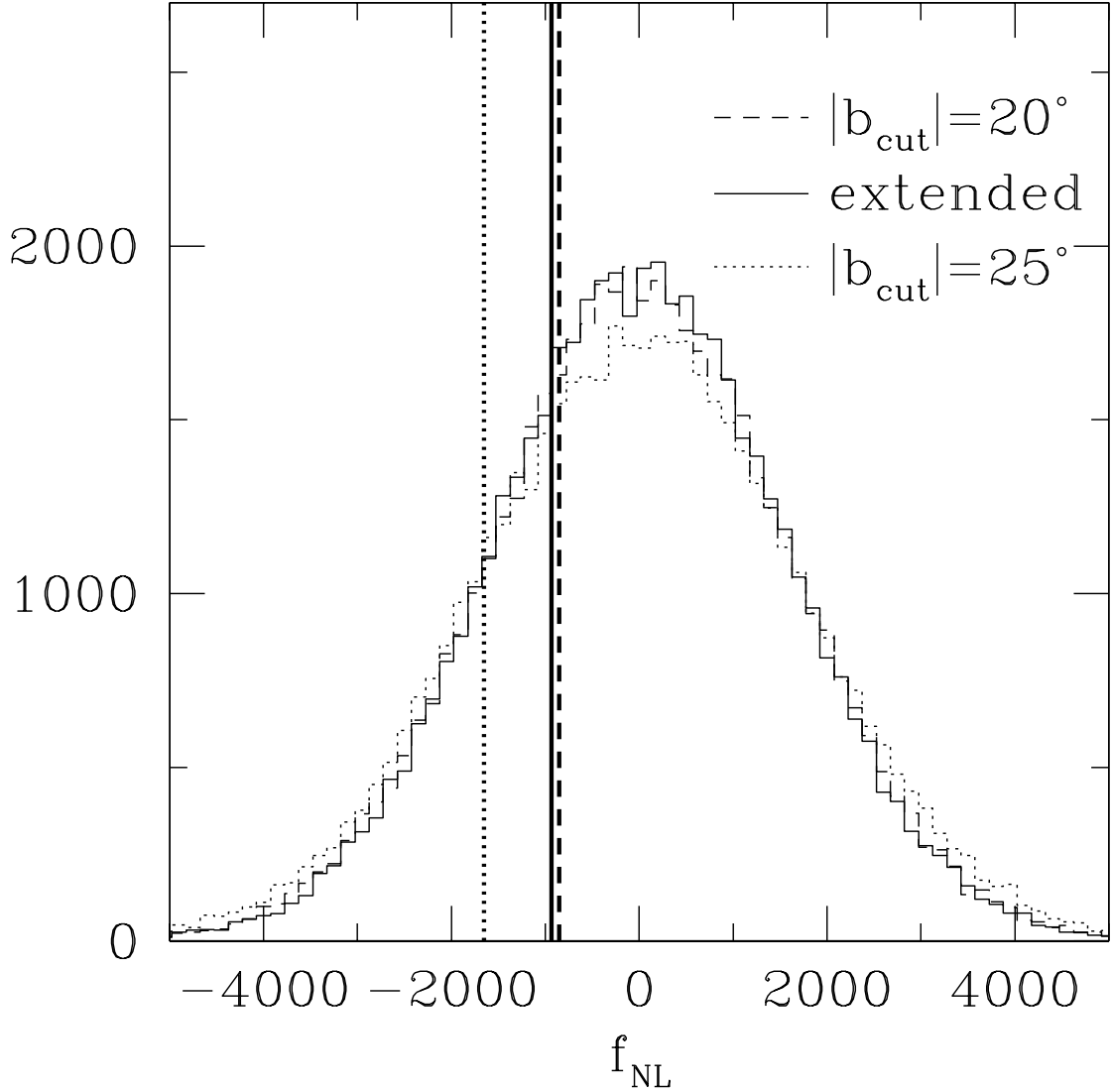}
 \caption
 {Constraint on the non-linear coupling parameter, $f_{\rm NL}$, which
 characterizes non-linearity in inflation (Eq.(\ref{eq:modelreal})).
 The dashed, solid, and dotted lines represent the three different 
 Galactic cuts as quoted in the figure.
 The thick vertical lines plot the measured values of $f_{\rm NL}$
 from the {\it COBE} DMR maps, while the histograms plot those drawn
 from the Monte--Carlo simulations for each cut.
 68\% confidence limits on $f_{\rm NL}$ are 
 $\left|f_{\rm NL}\right|<1.6\times 10^3$, $1.5\times 10^3$, and
 $1.7\times 10^3$ for the three cuts, respectively.}
\label{fig:fNL}
\end{figure}

Figure~\ref{fig:fNL} plots the measured values of the non-linear
coupling parameter, $f_{\rm NL}$, as well as the simulated realizations, 
for the three different Galactic cuts.
The measured values are well within the cosmic variance: we place 
68\% confidence limits on $f_{\rm NL}$ as
$\left|f_{\rm NL}\right|<1.6\times 10^3$, $1.5\times 10^3$, and
$1.7\times 10^3$, for the $20^\circ$ cut, the extended cut,
and the $25^\circ$ cut, respectively.

\begin{figure}
 \plotone{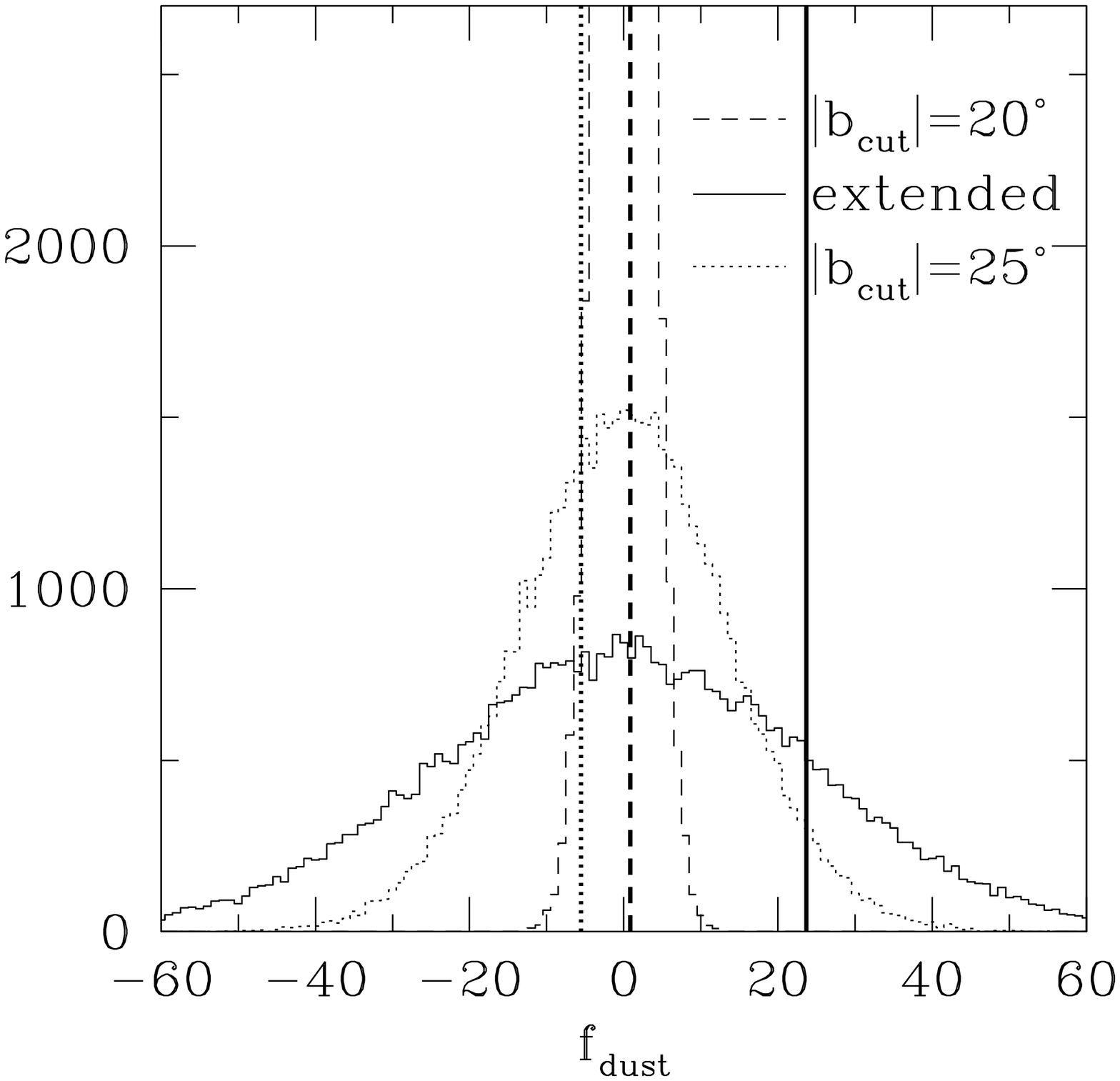}
 \caption
 {Constraint on amplitude of the interstellar dust bispectrum, 
 $f_{\rm dust}$.
 The meaning of the lines is the same as in figure~\ref{fig:fNL}.}
\label{fig:fdust}
\end{figure}

\begin{figure}
 \plotone{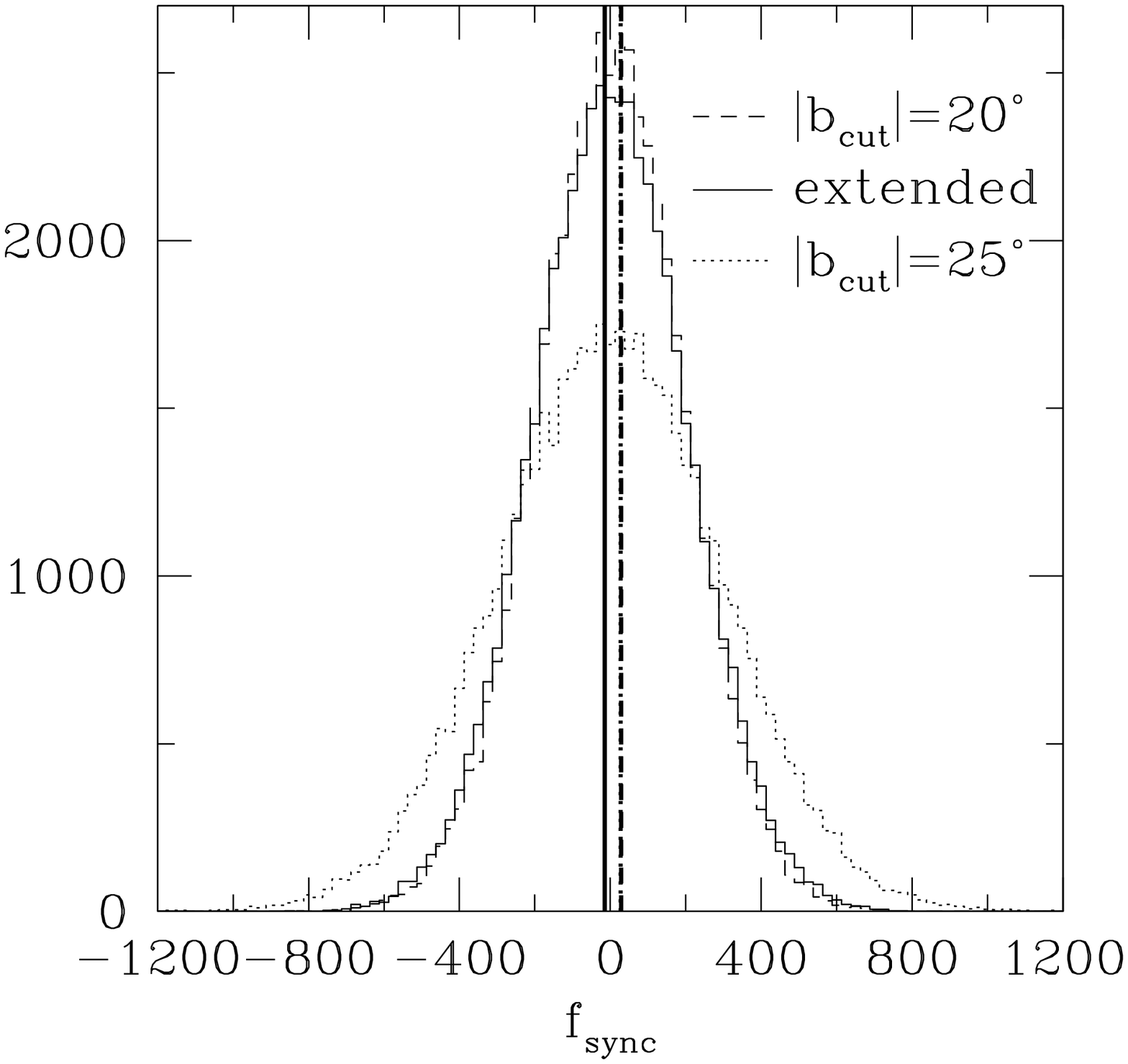}
 \caption
 {Constraint on amplitude of the interstellar synchrotron bispectrum,
 $f_{\rm sync}$.
 The meaning of the lines is the same as in figure~\ref{fig:fNL}.}
\label{fig:fsync}
\end{figure}

Figures~\ref{fig:fdust} and \ref{fig:fsync} plot
constraints on $f_{\rm dust}$ and $f_{\rm sync}$, respectively.
There is no indication of either component contributing
to the measured bispectrum significantly.

\subsection{Null test of the normalized bispectrum}

Using $\chi^2$ defined by equation~(\ref{eq:chisq}), we can test
Gaussianity of the DMR map. 
While the minimization of $\chi^2(f_j)$ gives constraints on 
the parameters, a value of $\chi^2(f_j)$ tells us goodness-of-fit;
$\chi^2(0)$ tests a hypothesis of the bispectrum being zero.
When $\chi^2(0)$ is either significantly greater or smaller than 
those drawn from the simulations, we conclude that the 
DMR map is inconsistent with zero bispectrum.

$\chi^2(0)$ is similar to what several authors have used
for quantifying statistical significance of non-Gaussianity in 
the DMR map \citep{FMG98,Mag00,SM00}.
They use only diagonal terms of the covariance matrix; however,
the matrix is diagonal only on the full sky.
As lack of sky coverage correlates one mode to the others, 
we should include off-diagonal terms as well.
We did so in equation~(\ref{eq:chisq}).

\begin{figure}
 \plotone{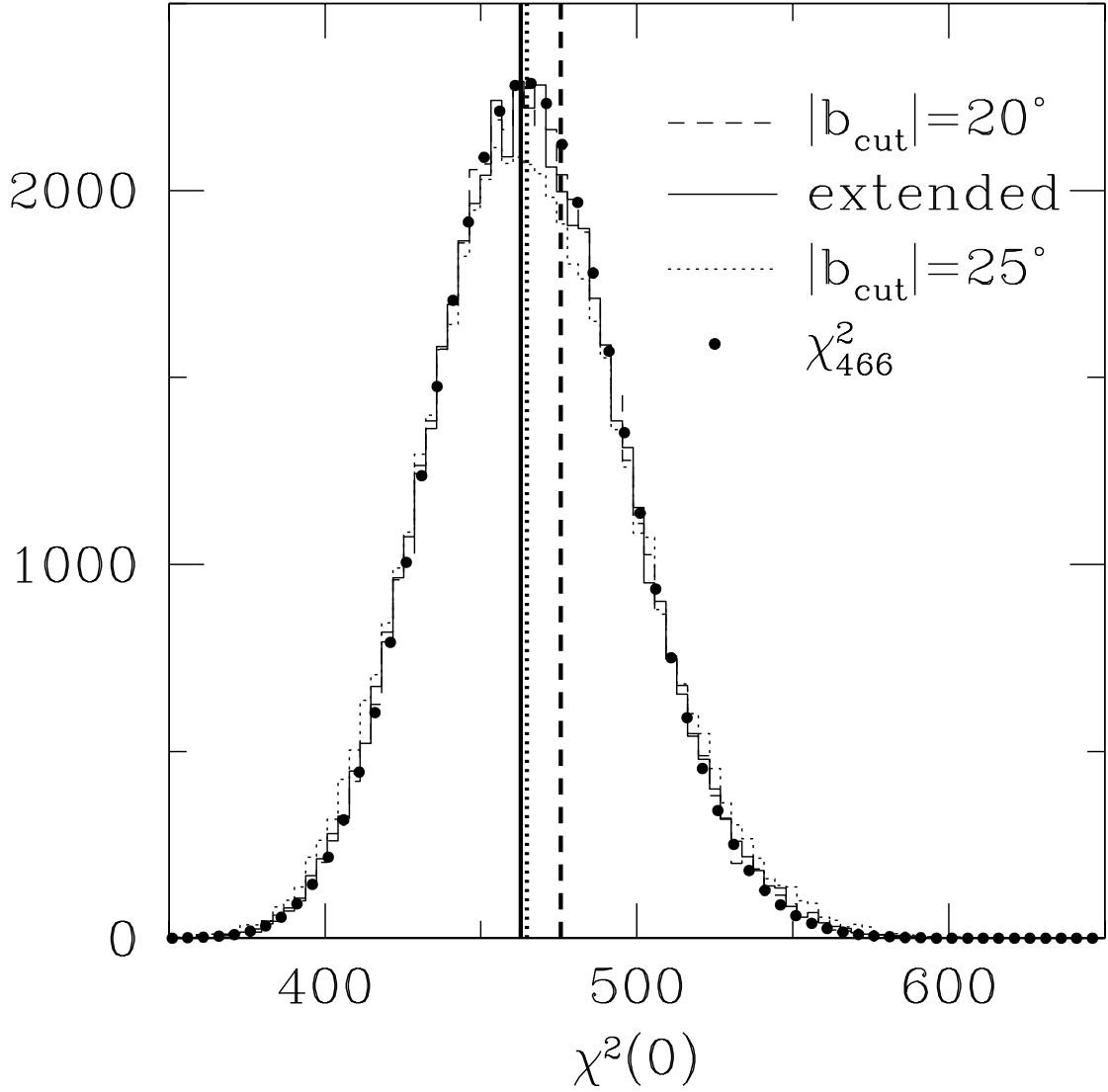}
 \caption
 {Testing hypothesis of the normalized bispectrum, 
 $B_{l_1l_2l_3}/\left(C_{l_1}C_{l_2}C_{l_3}\right)^{1/2}$, being zero
 in the {\it COBE} DMR four-year $53+90$~GHz sky map.
 The dashed, solid, and dotted lines represent 
 the $20^\circ$ cut, the extended cut, and 
 the $25^\circ$ cut, respectively.
 The thick vertical lines plot the measured $\chi^2(0)$, while the 
 histograms plot those drawn from the Monte--Carlo simulations.
 The filled circles plot the $\chi^2$ distribution for 466 degrees of 
 freedom.}
\label{fig:null_norm}
\end{figure}

Figure~\ref{fig:null_norm} compares $\chi^2_{\rm DMR}(0)$ with
$\chi^2_{\rm MC}(0)$ for the different Galactic cuts.
The measured values are $\chi^2_{\rm DMR}(0)=475.6$, 462.7, and 
464.8 for the corresponding Galactic cuts, respectively, while 
$\left<\chi^2_{\rm MC}(0)\right>=466$.
We find the probability of $\chi^2_{\rm MC}(0)$ being larger than 
$\chi^2_{\rm DMR}(0)$ to be 
$P\left(\chi^2_{\rm MC}>\chi^2_{\rm DMR}\right)=36.9\%$, 47.0\%, and 
49.7\%, respectively.

We conclude that the DMR map is comfortably consistent with zero 
normalized bispectrum.
We explain the claimed detection \citep{FMG98} by a statistical 
fluctuation as an alternative to the "eclipse effect" proposition 
made in \citet{BZG00}.

There is no evidence that the scatter of
the normalized bispectrum is too small to be consistent with
Gaussian, in contrast to the claim of \citet{Mag00} based on 
$\chi^2(0)$ derived from 8 modes.
To clarify, our analysis does not reject the possibility that the CMB sky
is non-Gaussian for only a small number of modes; however, in the absence
of a theoretical motivation for limiting the analysis to a specific set of
modes, we choose to treat all the bispectrum modes on an equal footing.
\citet{SM00} claim that the non-Gaussianity found by \cite{Mag00}
does not spread to other modes. 
This is consistent with our result.

Incidentally, we plot in the figure~\ref{fig:null_norm} 
the $\chi^2$ distribution for 466 degrees of freedom, 
$\chi^2_{466}$, in filled circles; we find that the distribution of 
$\chi^2_{\rm MC}(0)$ is very similar to the $\chi^2_{466}$ distribution 
for a smaller cut as expected, while it becomes 
slightly broader for a larger cut for which the distribution of the 
normalized bispectrum deviates from Gaussian appreciably. 
Yet, we find that the $20^\circ-25^\circ$ cuts reasonably 
retain the Gaussianity of the distribution of the normalized bispectrum.

\section{DISCUSSION AND CONCLUSIONS}
\label{sec:discussion_bl}

In this paper, we have measured all independent 
configurations of the angular bispectrum on the {\it COBE} DMR map,
down to the DMR beam size.
Using the most sensitive sky map to CMB, which combines the maps at 
53 and 90~GHz, we test the Gaussianity of the DMR map.

We find that the normalized bispectrum, 
$B_{l_1l_2l_3}/\left(C_{l_1}C_{l_2}C_{l_3}\right)^{1/2}$, 
gives more robust test of Gaussianity than the bare bispectrum, 
$B_{l_1l_2l_3}$. 
We compare the measured data with the simulated realizations,
finding the DMR map comfortably consistent with Gaussian.
We explain the reported detection of the normalized bispectrum
at $l_1=l_2=l_3=16$ \citep{FMG98} by a statistical fluctuation.
While it is still conceivable that the eclipse effect of the Earth against 
the {\it COBE} satellite generates some of the bispectrum \citep{BZG00},
the DMR data cannot distinguish it from the statistical fluctuations.

We fit the predicted bispectra to the data, constraining the parameters 
in the predictions, which include the primary bispectrum from inflation
and the foreground bispectra from interstellar dust and synchrotron
emissions.
We find that neither dust nor synchrotron emissions contribute to
the bispectrum significantly.

We have obtained a weak constraint on the non-linear coupling parameter,
$f_{\rm NL}$, that characterizes non-linearity in inflation. 
We interpret the constraint in terms of a single field inflation as follows. 
According to the analysis of non-linear perturbations on super horizon 
scales \citep{SB90}, we can explicitly calculate $f_{\rm NL}$ as
\begin{equation}
 \label{eq:SB91}
  f_{\rm NL}= 
  -\frac{5}{24\pi G}
  \left(\frac{\partial^2\ln H}{\partial\phi^2}\right),
\end{equation}
where $H$ is the Hubble parameter during inflation.
When applying the slow-roll conditions to an inflaton potential 
$V(\phi)$, we have $\partial\ln H/\partial\phi\approx (d\ln V/d\phi)/2$;
thus, $f_{\rm NL}$ is on the order of curvature of a slow-roll potential,
implying that $\left|f_{\rm NL}\right|$ should be smaller than 1 in
slow-roll inflation.
Therefore, the obtained constraint, 
$\left|f_{\rm NL}\right|<1.5\times 10^3$, seems too weak to be 
interesting; however, any deviation from slow-roll could yield larger 
$\left|f_{\rm NL}\right|$, bigger non-Gaussianity.

The next generation satellite experiments, the {\it Microwave Anisotropy
Probe} ({\it MAP}) and {\it Planck}, should be able to put more
stringent constraints on $f_{\rm NL}$. 
\citet{KS01} have shown 
that {\it MAP} and {\it Planck} should be sensitive down to 
$\left|f_{\rm NL}\right|\sim 20$ and 5, respectively.
We find that the actual constraint from {\it COBE} (figure~\ref{fig:fNL}) is 
much worse than the estimate.
This is partly due to different cosmology used for the model, but mainly 
due to incomplete sky coverage; the statistical power of the 
bispectrum at low multipoles is significantly weakened by the Galactic cut.
Since {\it MAP} and {\it Planck} probe much smaller angular scales, 
and their better angular resolution makes an extent of the Galactic cut 
smaller,
the degradation of sensitivity should be minimal. 
Moreover, the improved frequency coverage of future experiments 
will aid in extracting more usable CMB pixels from the data.
At this level of sensitivity, any deviation from slow-roll could give 
an interesting amount of the bispectrum, and {\it MAP} and {\it Planck}
will put severe constraints on any substantial deviation from slow-roll.

While we have explored adiabatic generation of the 
bispectrum only, isocurvature perturbations from inflation also generate
non-Gaussianity \citep{LM97,P97,BZ97}. 
They are in general more non-Gaussian than the adiabatic
perturbations; it is worth constraining those models by the same
strategy as we have done in this paper.

\acknowledgments

We would like to thank Charles L. Bennett, Gary Hinshaw, Misao Sasaki, 
Licia Verde, and Edward L. Wright for helpful discussions.
We would like to thank Uro$\check{\rm s}$ Seljak and Matias Zaldarriaga 
for making their {\sf CMBFAST} code publicly available.
E. K. acknowledges financial support from the Japan Society for
the Promotion of Sciences.
D. N. S. and B. D. W. are partially supported by the MAP/MIDEX program.

\appendix
\section{ANGULAR POWER SPECTRUM AND BISPECTRUM ON THE INCOMPLETE SKY}

Incomplete sky coverage destroys orthonormality of the 
spherical harmonics on the sky.
The degree to which orthonormality is broken is often
characterized by the coupling integral \citep{P80},
\begin{equation}
 \label{eq:coupling}
  W_{ll'mm'}
  \equiv
  \int
  d^2\hat{\mathbf n}~
  W(\hat{\mathbf n})
  Y_{lm}^*\left(\hat{\mathbf n}\right)
  Y_{l'm'}\left(\hat{\mathbf n}\right)
  =
  \int_{\Omega_{\rm obs}} 
  d^2\hat{\mathbf n}~
  Y_{lm}^*\left(\hat{\mathbf n}\right)
  Y_{l'm'}\left(\hat{\mathbf n}\right),
\end{equation}
where $W(\hat{\mathbf n})$ is zero in a cut region otherwise 1, and 
$\Omega_{\rm obs}$ denotes a solid angle of the observed sky.
When $W_{ll'mm'}\neq \delta_{ll'}\delta_{mm'}$,
the measured harmonic transform of the temperature 
anisotropy field, $a_{lm}$, becomes a 
{\it biased} estimator of the true harmonic transform, 
$a_{lm}^{\rm true}$, through
\begin{equation}
 \label{eq:alm_obs}
  a_{lm}=
  \sum_{l'=0}^\infty\sum_{m'=-l'}^{l'}a_{l'm'}^{\rm true} W_{ll'mm'}.
\end{equation}
Hence, we must correct our estimators of the power spectrum 
and the bispectrum for the bias arising from incomplete sky coverage.

First, we derive a relationship between the angular power spectrum
on the incomplete sky and that on the full sky. 
Taking the ensemble average of the estimator of the power spectrum,
the pseudo-$C_l$ \citep{WHG98,WHG00},
$C_l=(2l+1)^{-1}\sum_m\left|a_{lm}\right|^2$,
we have
\begin{eqnarray}
 \nonumber
 \left<C_l\right>
  &=& \frac1{2l+1}\sum_{l'}
  C_{l'}^{\rm true}
  \sum_{mm'}
  \left|W_{ll'mm'}\right|^2\\
 \nonumber
  &\approx&
  \frac1{2l+1}
  C_{l}^{\rm true}
  \sum_{m}
  \sum_{l'm'}
  \int 
  {d^2\hat{\mathbf n}}~
  W(\hat{\mathbf n})
  Y_{lm}^*\left(\hat{\mathbf n}\right)
  Y_{l'm'}\left(\hat{\mathbf n}\right)
    \int 
  {d^2\hat{\mathbf m}}~
  W(\hat{\mathbf m})
  Y_{lm}\left(\hat{\mathbf m}\right)
  Y_{l'm'}^*\left(\hat{\mathbf m}\right)\\
 \nonumber
  &=&
  \frac1{2l+1}
  C_{l}^{\rm true}
  \sum_m
  \int 
  {d^2\hat{\mathbf n}}~
  W(\hat{\mathbf n})
  Y_{lm}^*\left(\hat{\mathbf n}\right)
  \int 
  {d^2\hat{\mathbf m}}~
  W(\hat{\mathbf m})
  Y_{lm}\left(\hat{\mathbf m}\right)
  \delta^{(2)}\left(\hat{\mathbf n}-\hat{\mathbf m}\right)\\
 \nonumber
  &=&
  C_{l}^{\rm true}
  \int
  \frac{d^2\hat{\mathbf n}}{4\pi}~
  W(\hat{\mathbf n})
  P_{l}\left(1\right)\\
 &=&
  C_{l}^{\rm true}\frac{\Omega_{\rm obs}}{4\pi}.
\end{eqnarray}
In the second equality, we have taken $C_{l'}^{\rm true}$ out of the 
summation over $l'$,
as $\left|W_{ll'mm'}\right|^2$ peaks very sharply at $l=l'$, and
$C_{l'}^{\rm true}$ varies much more slowly than 
$\left|W_{ll'mm'}\right|^2$ in $l'$.
This approximation is good for nearly full sky coverage.
In the third equality, we have used 
$\sum_{l'm'}Y_{l'm'}\left(\hat{\mathbf n}\right)
Y_{l'm'}^*\left(\hat{\mathbf m}\right)=
\delta^{(2)}\left(\hat{\mathbf n}-\hat{\mathbf m}\right)$.
In the forth equality, we have used 
$\sum_m Y^*_{lm}\left(\hat{\mathbf n}\right)
Y_{lm}\left(\hat{\mathbf m}\right)= 
\frac{2l+1}{4\pi}P_l(\hat{\mathbf n}\cdot\hat{\mathbf m})$.
The result indicates that the bias amounts approximately to a 
fraction of the sky covered by observations.

Next, we derive a relationship between the angular bispectrum on the 
incomplete sky and that on the full sky. 
We begin with 
\begin{equation}
  \left<a_{l_1m_1}a_{l_2m_2}a_{l_3m_3}\right>
 = 
 \sum_{{\rm all}~l'm'}
 \left<a^{\rm true}_{l_1'm_1'}a^{\rm true}_{l_2'm_2'}
  a^{\rm true}_{l_3'm_3'}\right>
  W_{l_1l_1'm_1m_1'}W_{l_2l_2'm_2m_2'}W_{l_3l_3'm_3m_3'}.
\end{equation}
Rotational and parity invariance of the bispectrum 
implies the bispectrum given by
\begin{equation}
  \left<a_{l_1m_1}a_{l_2m_2}a_{l_3m_3}\right>
   =
   b_{l_1l_2l_3}
   \int
  d^2\hat{\mathbf n}~
  Y_{l_1m_1}^*\left(\hat{\mathbf n}\right)
  Y_{l_2m_2}^*\left(\hat{\mathbf n}\right)
  Y_{l_3m_3}^*\left(\hat{\mathbf n}\right),
\end{equation}
where $b_{l_1l_2l_3}$ is an arbitrary real symmetric function,
which is related to the angular averaged bispectrum, $B_{l_1l_2l_3}$.
When $b^{\rm true}_{l_1l_2l_3}$ varies much more slowly than 
the coupling integral, we obtain
\begin{eqnarray}
 \left<a_{l_1m_1}a_{l_2m_2}a_{l_3m_3}\right>
 &=&
  \nonumber
  \sum_{{\rm all}~l'}
  b_{l_1'l_2'l_3'}^{\rm true}
  \sum_{{\rm all}~m'}
  \int
  d^2\hat{\mathbf n}~
  Y_{l'_1m'_1}^*\left(\hat{\mathbf n}\right)
  Y_{l'_2m'_2}^*\left(\hat{\mathbf n}\right)
  Y_{l'_3m'_3}^*\left(\hat{\mathbf n}\right) \\
 \nonumber
 & &\times
 \int  d^2\hat{\mathbf n}_1~
  W(\hat{\mathbf n}_1)
  Y_{l_1'm_1'}\left(\hat{\mathbf n}_1\right)
  Y_{l_1m_1}^*\left(\hat{\mathbf n}_1\right) \\
 \nonumber
 & &\times
  \int  d^2\hat{\mathbf n}_2~
  W(\hat{\mathbf n}_2)
  Y_{l_2'm_2'}\left(\hat{\mathbf n}_2\right)
  Y_{l_2m_2}^*\left(\hat{\mathbf n}_2\right) \\
 \nonumber
 & &\times
  \int  d^2\hat{\mathbf n}_3~
  W(\hat{\mathbf n}_3)
  Y_{l_3'm_3'}\left(\hat{\mathbf n}_3\right)
  Y_{l_3m_3}^*\left(\hat{\mathbf n}_3\right)\\
 \label{eq:bl_obs}
  &\approx&
   b_{l_1l_2l_3}^{\rm true}
   \int  d^2\hat{\mathbf n}~
  W(\hat{\mathbf n})
  Y_{l_1m_1}^*\left(\hat{\mathbf n}\right)
  Y_{l_2m_2}^*\left(\hat{\mathbf n}\right)
  Y_{l_3m_3}^*\left(\hat{\mathbf n}\right).
\end{eqnarray}
Then, we calculate the angular averaged bispectrum, $B_{l_1l_2l_3}$
(Eq.(\ref{eq:blll})). 
By convolving equation~(\ref{eq:bl_obs}) with the Wigner-3$j$ symbol
and using the identity~(\ref{eq:gaunt}),
we obtain
\begin{eqnarray}
 \nonumber
 \left<B_{l_1l_2l_3}\right>
 &\approx&
 b_{l_1l_2l_3}^{\rm true}
 \sqrt{\frac{4\pi}{(2l_1+1)(2l_2+1)(2l_3+1)}}
   \left(\begin{array}{ccc}l_1&l_2&l_3\\0&0&0\end{array}\right)^{-1}\\
 \nonumber
  & &\times
 \sum_{{\rm all}~m}
  \int
  d^2\hat{\mathbf m}~
  Y_{l_1m_1}\left(\hat{\mathbf m}\right)
  Y_{l_2m_2}\left(\hat{\mathbf m}\right)
  Y_{l_3m_3}\left(\hat{\mathbf m}\right) \\
 & &\times 
 \nonumber
  \int
  d^2\hat{\mathbf n}~
  W(\hat{\mathbf n})
  Y_{l_1m_1}^*\left(\hat{\mathbf n}\right)
  Y_{l_2m_2}^*\left(\hat{\mathbf n}\right)
  Y_{l_3m_3}^*\left(\hat{\mathbf n}\right)\\
 \nonumber
 &=&
 b_{l_1l_2l_3}^{\rm true}
 \sqrt{\frac{(2l_1+1)(2l_2+1)(2l_3+1)}{4\pi}}
   \left(\begin{array}{ccc}l_1&l_2&l_3\\0&0&0\end{array}\right)^{-1}\\
 \nonumber
  & &\times
  \int  \frac{d^2\hat{\mathbf m}}{4\pi}
  \int  \frac{d^2\hat{\mathbf n}}{4\pi}~
  W(\hat{\mathbf n})
  P_{l_1}\left(\hat{\mathbf m}\cdot\hat{\mathbf n}\right)
  P_{l_2}\left(\hat{\mathbf m}\cdot\hat{\mathbf n}\right)
  P_{l_3}\left(\hat{\mathbf m}\cdot\hat{\mathbf n}\right)\\
 \nonumber
 &=&
  b_{l_1l_2l_3}^{\rm true}
  \sqrt{\frac{(2l_1+1)(2l_2+1)(2l_3+1)}{4\pi}}
	\left(\begin{array}{ccc}l_1&l_2&l_3\\0&0&0\end{array}\right)
	\frac{\Omega_{\rm obs}}{4\pi}\\
 &=&
  B_{l_1l_2l_3}^{\rm true}\frac{\Omega_{\rm obs}}{4\pi},
\end{eqnarray}
where we have used the identity,
\begin{equation}
 \int_{-1}^{1}\frac{dx}2~
  P_{l_1}(x)P_{l_2}(x)P_{l_3}(x)
  =
  \left(\begin{array}{ccc}l_1&l_2&l_3\\0&0&0\end{array}\right)^2.
\end{equation}
Thus, the bias for the angular bispectrum on the incomplete sky is also 
approximately given by a fraction of the sky covered by observations.


\end{document}